\documentclass{article}
\usepackage{graphicx,epsfig}
\newcommand{\re}[1]{(\ref{#1})}

\begin{document}
\pagestyle{empty}
\begin{flushright}
TTP98-39\\
PM/98-35\\
hep-ph/9811275\\
\end{flushright}
\vspace*{4mm}
\begin{center}
{\Large\bf{
Short-distance tachyonic gluon mass and $1/Q^2$ corrections
}} \\
\vspace{1.5cm}

{\bf K.G. Chetyrkin}\footnote{On leave from Institute for Nuclear Research
of the Russian Academy of Sciences, Moscow, 117312, Russia.},\\

\vspace{0.3cm}
Institut f\"ur Theoretische Teilchenphysik, Universit\"at Karlsruhe, \\
Kaiserstrasse 12, 76128 Karlsruhe, Germany.\\ E-mail:
chet@particle.physik.uni-karlsruhe.de \\ \vspace{0.5cm} {\bf Stephan
Narison}
\\
\vspace{0.3cm}
Laboratoire de Physique Math\'ematique et Th\'{e}orique\\ UM2, Place
Eug\`ene Bataillon\\
34095 Montpellier Cedex 05, France\\
E-mail:
narison@lpm.univ-montp2.fr\\
\vspace{0.5cm}
{\bf V.I. Zakharov}\\
\vspace{0.3cm}
Max-Planck Institut f\"ur Physik,\\
F\"ohringer Ring 6, 80805 M\"unchen, Germany.\\ E-mail: xxz@mppmu.mpg.de \\
\vspace{1.5cm}

\begin{abstract}

\noindent
We consider the assumption that a tachyonic gluon mass imitates
short-distance nonperturbative physics of QCD. The phenomenological
implications include modifications of the QCD sum rules for correlators of
currents with various quantum
numbers. The new $1/Q^2$ terms allow to resolve in a natural way old
puzzles in the pion and
scalar-gluonium channels. They lead
to a slight reduction of the values of the running light quark masses from
the (pseudo)scalar sum rules and of $\alpha_s(M_\tau)$ from $\tau$ decay
data. Analogously such terms only affect slightly the determinations of the
running strange quark mass from $e^+e^-$ and $\tau$ decay data.
Further tests can be provided by precision measurements of the correlators on
the lattice and by the $e^+e^-\rightarrow$ hadrons data. \end{abstract}
\end{center}

\vfill\eject
\pagestyle{plain}
\setcounter{page}{1}
\section{Introduction}
In this paper we will outline phenomenology of nonperturbative
short-distance physics in QCD
based on introduction of a tachyonic gluon mass. \\ To set up a theoretical
framework, we
will consider physical observables characterized by a large mass scale $Q$,
$Q^2>> \Lambda_{QCD}^2$
where $\Lambda_{QCD}$ is the position of the Landau pole in the running
coupling:
\begin{equation}
\alpha_s(Q^2)\simeq \frac{1}{b_0\ln{\left(
{Q^2}/{\Lambda_{QCD}^2}\right)}}, \end{equation}
where: \begin{equation}\label{beta}
b_0\equiv -\frac{\beta_1}{2\pi}=\frac{1}{4\pi}\left( 11-\frac{2}{3}n_f\right)
\end{equation}
for $n_f$ flavours, is the first coefficient of the $\beta$ function. Then
in the limit of $Q^2\rightarrow\infty$, QCD reduces to the parton model
while at finite $Q^2$ there
are corrections of order $(1/\ln Q)^k$ which are nothing else but
perturbative expansions and power corrections, $(\Lambda_{QCD}/Q)^n$ which
are
nonperturbative in nature.
\\
To be more specific, concentrate on
the correlators of currents $J$
with various quantum numbers:
\begin{equation}
\Pi_J (Q^2)~=~i\int e^{iqx}\langle 0|T\{J(x),J(0)\}|0\rangle d x \,
,\label{corr}
\end{equation}
where $Q^2\equiv-q^2$ and we suppress for the moment the Lorentz indices.
Furthermore, to stick
to the standard notations \cite{svz,snb} we introduce the Borel/ Laplace
transformed $\Pi(M^2)$ where:
\begin{equation}
\Pi_J (M^2)~\equiv~{ Q^{2n}\over (n-1)!}\left({-d\over dQ^2}\right)^n\Pi_J
(Q^2) \end{equation}
in the limit where both $Q^2$ and $n$ tend to infinity so that their ratio
$M^2\equiv Q^2/n$ remains finite.
Then a standard representation for correlators (\ref{corr}) at large $Q^2$
goes back to
the QCD sum rules \cite{svz,snb} and somewhat schematically reads as:
\begin{equation}
\Pi_J(M^2)\approx (parton~model)\cdot
\left( 1+{a_J\over \ln\left( {M^2}/{\Lambda_{QCD}^2}\right)}+{c_J\over M^4}
+{\cal O}\left(
\ln^{-2}\left({M^2}/{\Lambda_{QCD}^2}\right), M^{-6}\right)\right)
\label{st}\end{equation} where the constants $a_J,c_J$ depend on the
channel, i.e. on the quantum numbers of the current $J(x)$, and we assumed
that the currents have zero anomalous dimensions. Moreover the terms of
order $1/\ln M^2$ and $M^{-4}$ are associated with the first perturbative
correction and the gluon
condensate, respectively.
\\
One of the central points about Eq. (\ref{st}) is the absence of $M^{-2}$
corrections. The reason is that Eq. (\ref{st}) utilises the standard
operator product expansion (OPE) and there are no gauge invariant operators
of dimension $d=2$ which could have vacuum-to-vacuum matrix elements. \\ We
will consider a modified phenomenology accounting for a correction due to a
(postulated)
non-vanishing gluon mass squared, $\lambda^2$. In this approach Eqs
(\ref{st}) are replaced by
\begin{equation}
\Pi_j(M^2)~\approx~(parton~model)\cdot\left( 1+{a_J\over \ln\left(
{M^2}/{\Lambda_{QCD}^2}\right)}+{b_J\over M^2}+{c_J\over M^4} +{\cal
O}\left( \ln^{-2}\left({M^2}/{\Lambda_{QCD}^2} \right),
M^{-6}\right)\right)\label{b}
\end{equation} where $b_J$ are calculable in terms of $\lambda^2$ and
depend on the
channel considered. As we
shall explain later, the introduction of the gluon mass is a heuristic way
to account for possible
existence of small-size strings. Moreover, it is easy to realize that in
this case we expect a
tachyonic gluon mass, $\lambda^2< 0$. Also, we have to confine ourselves to
the cases when the
$\lambda^2$ correction is associated with short distances. \section{Power
corrections beyond the OPE} The validity of the OPE beyond the perturbation
theory cannot be proven without identifying more precisely nonperturbative
mechanisms. One may discuss both infrared- and ultraviolet-sensitive
nonperturbative contributions. As was emphasised rather recently
\cite{grunberg} the general arguments based on dispersion relations alone
cannot rule out extra $1/Q^2$ pieces. One can substantiate the point by the
simple observation that the removal of the Landau pole from the running
coupling:
\begin{equation}\label{alfauncov}
{\alpha_s}(Q^2)\simeq {1\over b_0\ln{\left( Q^2/\Lambda_{QCD}^2\right)}}~
\rightarrow~{1\over b_0\ln{\left( Q^2/\Lambda_{QCD}^2\right)}}-
{\Lambda_{QCD}^2\over
b_0(Q^2-\Lambda_{QCD}^2)}\end{equation}
does introduce
a $\Lambda_{QCD}^2/Q^2$ term at large $Q^2$. Note that the ad hoc
modification of the coupling in the IR region cannot be accommodated by the
OPE. This simple example emphasises that the standard OPE is a dynamical
framework based on a particular analysis of a generic Feynman graph, with
large $Q$ flowing in and out, and allowing some of the lines to
become soft \cite{svz,snb}. In this paper we accept the standard OPE to
treat the infrared-sensitive contributions. \\ As for the
ultraviolet-sensitive contributions, an insight into power corrections is
provided by renormalons
(for review and further references see \cite{review}). Renormalons are a
particular set of perturbative graphs which result in a divergent
perturbative
series, $\sum_na_n\alpha_s^n (Q^2)$ with $a_n\sim n!(-b_0)^n$. Starting
with $n=N_{cr}\equiv
1/b_0\alpha_s(Q^2)$ the product $|a_n\alpha_s^n|$ grows with $n$ despite of
the smallness of
$\alpha_s(Q^2)$. The most common assumption is that the rising branch of
the perturbative expansion is
summed up a la Borel:
\begin{equation}
\sum_{N_{cr}}^{\infty}n!(-b_0)^n\alpha_s^n(Q^2)\rightarrow \int
dt~{(-b_0\alpha_s(Q^2)\cdot
t)^{N_{cr}}exp(-t)\over 1 +b_0\alpha_s(Q^2)\cdot t}\sim
{\Lambda_{QCD}^2\over Q^2}. \end{equation} Although some further arguments
can be given to support this assumption \cite{beneke1},
the Borel summation still looks an arbitrary solution and the theory is
basically undefined
on the level of $\Lambda_{QCD}^2/Q^2$ terms. \\
Hence, one may argue that there should be nonperturbative terms of the same
order, $\Lambda_{QCD}^2/Q^2$ which make the theory uniquely defined. Note
that this kind of logic is very common in case of infrared renormalons
(see, e.g., \cite{review}). There were a few attempts to develop a
phenomenology of unconventional $\Lambda_{QCD}^2/Q^2$ corrections
\cite{yamawaki}. Since there is no
means to evaluate the nonperturbative terms in various channels directly
the crucial ingredient here is the assumption on the nature of
nonperturbative terms. In particular, a chiral-invariant quark mass and
dominance of four-fermionic operators a la Nambu-Jona-Lasino have been
tried for phenomenology
\cite{yamawaki}.
\\
Here we will explore the phenomenology of the $1/Q^2$ corrections assuming
that at
short distances they can be effectively described by a tachyonic gluon
mass. The assumption is of openly heuristic nature and can be motivated in
the following way. Consider
the static potential $V(r)$ of heavy quarks and represent it as:
\begin{equation}
V(r)~\approx~-{4\alpha_s(r)\over 3r}+kr\label{potential} .\end{equation}
Such a form
of the potential is quite common at large distances $r$ with $k\approx 0.2$
GeV$^2$
representing the string tension. At {\it short} distances one expects that
the linear correction to the
Coulomb-like potential is replaced by a $r^2$ term \cite{balitsky}. The
vanishing of the linear
correction at short distances is a reflection of the absence of the
$Q^{-2}$ correction in the
current correlators (see above). And vice versa, if the coupling
$\alpha_s(Q^2)$ has an unconventional
$Q^{-2}$ piece then the potential $V(r)$ has a linear correction at short
distances as well
\cite{az2}.
%
%
\\
The power-like behaviour of the gluon propagator was a subject of intense
theoretical studies both by means of the lattice simulations (see,
e.g.,~\cite{Leinweber} and references therein) and by means of the
Schwinger-Dyson equation (see, e.g.,~\cite{Buttner:1995hg} and references
therein). However, these
studies refer to large distances. As for the short distances, the smallest
distances where the measurements of the potential on the lattice are
available are of order 0.1fm \cite{bali}. Remarkably, there is no sign of
the
vanishing of the $kr$ term so far.
%
%
Although it is not a proof of the
presence of the linear
term at $r\rightarrow 0$ since the measurements were not specifically
dedicated to short distances,
such a possibility cannot at all be ruled out and one is free to speculate
that the linear term does
not vanish at short distances. Thus, let us assume that the form
(\ref{potential}) remains indeed true
as $r\rightarrow 0$. There is still no obvious way to evaluate $1/Q^2$
corrections to other variables
in terms of the same $k$ . Note, however, that as far as the $kr$ term is
treated as a small
correction at short distances, it can be reproduced by a tachyonic gluon
mass $\lambda^2$ where
\cite{vz2} \begin{equation}
-{2 \alpha_s\over 3}\lambda^2~\approx~ k\label{mass} {}.
\end{equation}
In
this framework $\lambda^2$
is intended to represent the effect of small-size strings. From a
calculational point of view, we simply replace the gluon propagator
\begin{equation}
D_{\mu\nu}^{ab}(k^2)~=~{\delta^{ab}\delta_{\mu\nu}\over k^2}~\rightarrow~
\delta^{ab}\delta_{\mu\nu}\left({1\over k^2}+{\lambda^2\over k^4}
\right)\label{simple}
\end{equation}
and check that the integral is saturated by large momenta, $k^2\sim Q^2$.
One should notice that, to
the approximation we are working, our analysis is gauge invariant. (An
attempt to
generalise the substitution in Eq. (\ref{simple}) to higher loops can be
found in Ref.
\cite{anishetty}).
\\
There are obvious limitations to this approach. In particular, we do not
include terms of order $\lambda^4$ and do not go beyond one loop. Moreover,
to use the Eq. (\ref{simple}) consistently at all distances one should have
evaluated the anomalous dimension of the string tension $k$ which goes
beyond the scope of the present note
(see, however, \cite{anishetty}). Also, the effect of strings at large
distances cannot be reproduced by any modified propagator \cite{gubarev}
and this is one more reason to stick to short distances. Finally, the
procedure described above assumes that at short distances the $kr$ piece in
the potential is associated with a vector exchange while at large distances
it corresponds to a scalar exchange and there is no direct evidence to
support such a change \cite{bali}.
\\
With these reservations in mind, it is clear that we would better
concentrate first of all on the
qualitative features of the approach with $\lambda^2\neq 0$. Let us
emphasise therefore from the very beginning that there are qualitative
effects around. Indeed,
the estimate in Eq. (\ref{mass}) implies a large numerical value for
$\lambda^2$. Applying
Eq. (\ref{mass}), for example, at distances corresponding to $\alpha_s=0.3$
we would find
$\lambda^2\approx -1 $ GeV$^2$. Although this estimate cannot be trusted to
the accuracy, say,
better than factor of 2 such a big gluon mass might seem to be easily ruled
out since it would
produce too much distortion on the standard phenomenology. We shall argue
that it is actually not
the case and $\lambda^2\approx -0.5 $ GeV$^2$ can well be admittable.
\section{Puzzles to be resolved}
Turning to the phenomenology,
let us first note that not all the $1/Q^2$ corrections in QCD are
associated with short distances. Respectively, not all the terms
proportional to $\lambda^2$ can be
used to mark short-distance contributions. For example, in case of DIS the
$1/Q^2$ corrections are
coming from the IR region and perfectly consistent with the OPE. Thus, the
class of theoretical
objects for which an observation of the $1/Q^2$ corrections would signify
breaking of the OPE is
limited. One example was the potential $V(r)$ discussed above. Other
examples are the correlator
functions (\ref{corr}) where the power corrections start with $Q^{-4}$
terms. In terms of the
expansion in $\lambda^2$, we are guaranteed then that terms proportional to
$\lambda^2$ are associated
with short distances. If such terms arose from the infrared region, then
the final result would
contain nonanalytical in $\lambda^2$ terms, $\lambda^2\ln\lambda^2$
\cite{braun} and we shall see that
it is not indeed the case.
\\
Even this limited set of variables exhibits actually a remarkable variety
of scales which are not explained by the standard OPE \cite{novikov}. We
will briefly review
these puzzles since it is an important question, whether a novel
phenomenology we are going to explore
can resolve the puzzles of the existing phenomenology. \\ One of the basic
quantities to be determined from the theory is at which scale the parton
model
predictions for the correlators in Eq. (\ref{corr}) get violated
considerably via the power
corrections. To quantify the scales inherent to various correlators in Eq.
(\ref{corr}) one may
introduce the notion of
$M^2_{crit}$ which is defined as the value of $M^2$ at which the power
corrections become 10 per cent from the
unit term \cite{novikov}. While
choosing just 10 per cent is a pure convention, the meaning of $M^2_{crit}$
is that at lower $M^2$ the power corrections blow up. Moreover, the values
of $M^2_{crit}$ can be evaluated using either experimental or theoretical
input. In the latter case the calculation is usually under control only as
far as the power correction is relatively small and that is why
$M^2_{crit}$ is chosen to refer to a 10 per cent, not 100 per cent
correction.

Consider
first the best studied $\rho$-channel. On the theoretical side,
$\Pi_{\rho}$ is represented as:
\begin{equation}
\Pi_{\rho}(M^2)~\approx~(parton~model)\cdot \left(1+{\alpha_s(M^2)\over
\pi}+ {\pi^2\over 3M^4}\langle{\alpha_s\over \pi}
(G^a_{\mu\nu})^2\rangle+...\right) \end{equation} The value of $M^2_{crit}$
is related
then to the value of the gluon
condensate, $\langle{\alpha_s\over \pi}(G^a_{\mu\nu})^2\rangle$:
\begin{equation} M^2_{crit}(\rho-channel)~\approx~\sqrt{10{\pi\over 3}
\langle\alpha_s G^2 \rangle}
~\approx~(0.6~-~0.8)~{\mbox{GeV}}^2\label{normal} \end{equation} where the
factor
of 10 in the r.h.s.
reflects the 10 per cent convention introduced above. An independent
information on $M^2_{crit}$ in
this channel can be obtained by using the dispersion relations for
$\Pi_{\rho}(M^2)$ and integrating over the corresponding experimental cross
section of
$e^+e^-$ annihilation into hadrons, or by optimising the $M^2$-dependence
of the Borel/Laplace
sum rules. The value of $M^2_{crit}$ is then most sensitive to parameters
of the
$\rho$-meson and comes
out about the same $0.6~{\mbox{GeV}}^2$ as indicated by Eq. (\ref{normal}).
Since the gluon
condensate parametrises contribution of the infrared region one may say
that at least in the
$\rho$-channel the breaking of the asymptotic freedom is due to infrared
phenomena. Recent
detailed experimental studies of the $\tau$-decays did not contradict this
picture
\cite{aleph}.
\\
If one proceeds to other channels, in particular to the $\pi$-channel and
to the
$0^{\pm}$-gluonium channels, nothing special happens to $M^2_{crit}$
associated with the
infrared-sensitive contribution parametrised by the gluon condensate.
However, it was determined from
independent arguments \cite{novikov} that the actual values of $M^2_{crit}$
do vary
considerably in these channels:
\begin{eqnarray} M^2_{crit}(\pi-channel)~\ge~ 2~{\mbox{GeV}}^2\\\label{pion}
M^2_{crit}(0^{\pm}-gluonium~channel)~\approx
15~{\mbox{GeV}}^2\label{gluonium} .\end{eqnarray} In the pion channel, the
lower
bound on $M^2_{crit}$ can be determined by evaluating the pion contribution
to the
dispersive integral and equating it to the parton-model result. Since all
other hadronic states
give positive-definite contributions, at lower $M^2$ the parton model is
certainly violated. In more
detail, one gets: \begin{equation}\label{pion1}
M^2_{crit}(\pi-channel)~\geq~\sqrt{{16\pi^2\over 3}
{f^2_{\pi}m_{\pi}^4\over (\bar{m}_u+\bar{m}_d)^2}}
\approx 1.8~\rm{GeV}^2,\end{equation}
where the running
light quark masses are evaluated
at the scale $M^2$, and $f_\pi$ = 93 MeV. A detailed $M^2$ optimisation
analysis of the pion sum
rule gives a value \cite{snb}:
\begin{equation}\label{pion2}
M^2_{crit}(\pi-channel)\approx (2\sim 2.5)~\rm{GeV}^2. \end{equation}
Although the
difference of factor (3-4) in
the values of $M^2_{crit}$ in the $\rho$- and $\pi$-channels might seem not
so big, it cannot be
matched by the standard IR contributions.

Note also
that the difference in
$M^2_{crit}$ in the $\rho$- and $\pi$-channels was confirmed by the direct
measurements of the corresponding
correlators on the lattice \cite{chu}.
These measurements also provide with an independent support to the idea
that asymptotic freedom is violated at moderate $M^2\sim 1~GeV^2$ sharply,
due to power corrections. It is most important in case of channels with
large perturbative corrections,
the fact which might give rise to doubts in relevance of the notion of
$M^2_{crit}$ \\ In case of the scalar-gluonium channel, the value of
$M^2_{crit}$ follows \cite{novikov} from a low-energy theorem for the
correlator associated with the current $\alpha_s(G^2_{\mu\nu})^2$. This
low-energy theorem can be translated into correction to the parton model
predictions at large $M^2$: \begin{equation} \Pi_G(M^2)\approx
(parton ~model)\cdot\left(1+\left(\frac{4}{b_0}\right)
\left(\frac{\pi}{\alpha_s}\right)^2
\frac{ \langle\alpha_s G^2 \rangle}{M^4} +...\right), \label{huge}
\end{equation}
where $b_0$ is defined in Eq. (\ref{beta}). Thus, the low-energy theorem
brings in a large numerical factor $12/(b_0\pi)\left(
\pi^2/\alpha_s^2\right)\approx 400$ in front of the $M^{-4}$ correction as
compared with the	$\rho$-channel. This changes dramatically the
estimate of
$M^2_{crit}$ to:
\begin{equation}\label{glue2}
M^2_{crit}(0^+-gluonium)~\approx~ 20M^2_{crit}
(\rho-channel)~\approx~15~\rm{GeV}^2~,
\end{equation}
and of the resonance properties, respectively (see Ref.\cite{nva,nv}), in
rough agreement
with
the $M^2$-stability
analysis of the scalar \cite{nva,nv} and pseudoscalar \cite{shore,nv}
gluonia sum rules. Moreover, the huge power correction (\ref{huge}) remains
of a uniquely large scale and is not supported by any other large-scale
correction which makes it very difficult to interpret the breaking of the
asymptotic freedom in terms of resonances \cite{novikov}. \\ Also, the
phenomenology build on the IR power corrections cannot resolve the problem
of
$\eta^{'}$-meson. As is known from general considerations \cite{thooft} a
dynamical resolution
of this problem is not possible without accounting for field configurations
of nontrivial
topology. On the other hand, the standard OPE is based on standard, i.e.
perturbative Feynman
graphs and cannot account for a nontrivial topology. Hence, one invokes
direct instantons when
the currents $J$ interact at large $Q^2$ with small-size instantons
\cite{novikov}. The
instanton-induced contributions to the correlators in Eq. (\ref{corr}) were
extensively studied in
the literature (for review and references see \cite{shuryak}). We shall further
comment on this point later.
\section{Some phenomenological implications} Now, we add a new term
proportional to $\lambda^2$ to the theoretical side of $\Pi_J(M^2)$, see
Eq.(\ref{b}). It is
convenient to consider the effect of $\lambda^2 \neq 0$ channel by channel.
\subsection{Constraint on $\lambda^2$ from the $\rho$-channel}
Phenomenologically, in the
$\rho$-channel there are severe restrictions \cite{narison} on the new term
$b_{\rho}$: \begin{equation}
b_{\rho}~\approx~-~(0.03-0.07)~{\rm GeV}^2\label{constr} .\end{equation} To
find out
the implications of
this constraint for the gluon mass, we turn to the correlator of two vector
currents
$J^\mu_V(x)\equiv \bar{\psi}_i\gamma^\mu\psi_j(x)$, viz. \begin{eqnarray}
\Pi_{V}(Q^2) &\equiv& i \int d^4x ~e^{iqx} \langle 0\vert {\cal T}
J^\mu_{V}(x)
\left( J^\nu _{V}\right)\dagger (0) \vert 0 \rangle ,\nonumber\\
&-&(g^{\mu\nu}q^2-q^\mu
q^\nu)\Pi^{(1)}_{V}(Q^2)
+ q^\mu q^\nu\Pi^{(0)}_{V}(Q^2).
\label{vec.corr.def}
\end{eqnarray}
Here the indexes $i,j$ correspond to quark flavours; $m_i$ is the mass of
the quark $i$.
To first order in $\alpha_s$ and expanding in $m_{i,j}$ we obtain:
\footnote{The results described in
Eqs.~(\ref{Pi1m2GGv1},\ref{mqq},\ref{pis5:res}, \ref{scalarGG:res})
below have been obtained with the help of program packages MATAD
\cite{MATAD} and MINCER \cite{MINCER} written in FORM \cite{FORM}.}
\begin{equation}
\label{Pi1full}
\Pi^{(1)}_{V} =
\Pi^{(1)}_{V,con} + \Pi^{(1)}_{V,NO}
{}\, ,
\end{equation}
where
\begin{equation}
\label{Pi1con}
\Pi^{(1)}_{V,con} = \frac{m_j\langle \bar{\psi_i} \psi_i \rangle + m_i \langle
\bar{\psi_j} \psi_j \rangle}{QQ^2} \end{equation} and \begin{eqnarray}
\lefteqn{(16\pi^2)\Pi^{(1)}_{V,NO} =
}
\nonumber\\
&{+}&
\left[
\frac{20}{3}
+6 \frac{m_{-}^2}{Q^2}
-6 \frac{m_{+}^2}{Q^2}
+4 l_{\mu Q}
+6 \frac{m_{-}^2}{Q^2}l_{\mu Q}
\right]
\nonumber\\
&{+}& \frac{\alpha_s}{\pi}
\left[
\frac{55}{3}
-16 \,\zeta(3)
+\frac{107}{2} \frac{m_{-}^2}{Q^2}
-24 \,\zeta(3)\frac{m_{-}^2}{Q^2}
-16 \frac{m_{+}^2}{Q^2}
\right. \nonumber \\ &{}& \left.
\phantom{+ \frac{\alpha_s}{\pi}}
+4 l_{\mu Q}
+22 \frac{m_{-}^2}{Q^2}l_{\mu Q}
-12 \frac{m_{+}^2}{Q^2}l_{\mu Q}
+6 \frac{m_{-}^2}{Q^2}l_{\mu Q}^2
\right]
\nonumber\\
&{+}& \frac{\alpha_s}{\pi}\frac{\lambda^2}{Q^2} \left[ -\frac{128}{3} +32
\,\zeta(3)
-\frac{46}{3} \frac{m_{-}^2}{Q^2}
+16 \,\zeta(3)\frac{m_{-}^2}{Q^2}
-18 \frac{m_{+}^2}{Q^2}
\right. \nonumber \\ &{}& \left.
\phantom{+ \frac{\alpha_s}{\pi}\frac{\lambda^2}{Q^2} } +6
\frac{m_{-}^2}{Q^2}l_{iQ}
-6 \frac{m_{+}^2}{Q^2}l_{iQ}
+6 \frac{m_{-}^2}{Q^2}l_{jQ}
-6 \frac{m_{+}^2}{Q^2}l_{jQ}
\right]
{}.
\label{Pi1m2GGv1}
\end{eqnarray}
The above result is in $\overline{\mbox{MS}}$ scheme and the notations are
as follows:
\[
m_{\pm}=m_i\pm m_j, \ \ \
l_{ \mu Q}= \log(\frac{\mu^2}{Q^2}),\ \ \ \mbox{\rm and} \ \ \ l_{ i Q}=
\log(\frac{m_i^2}{Q^2}).
\]
The normal-ordered quark condensates in Eq.~\re{Pi1full} do not receive by
definition
any perturbative corrections and are displayed explictly only to make
easier the discussion below. Note
that the terms of order $\lambda^2/Q^2$ in Eq.~(\ref{Pi1m2GGv1}) are $\mu$
independent and, thus, do
not depend on the way how the overall UV subtraction of the correlator in
Eq. (\ref{vec.corr.def}) has
been fixed. Eq.~(\ref{Pi1m2GGv1}) are written for the case of normal
ordered quark condensate, that is
the reason why quark mass logs appear there. As is well-known these are of
long-distance nature and can
(and even should) be absorbed into the quark condensate
\cite{BroadGen84,CheSpi88,Jamin:1993se}. To do this	we should
use normal non-ordered condensates, which means that perturbative theory
contributions should {\em not}
be automatically subtracted from the latter. Instead, these contributions
are to be minimally
renormalized.
\\
In order $\alpha_s$ there are two diagrams (see Fig. 1) leading to nonzero
vacuum expectation value of
$
\langle \bar{\psi_i} \psi_i \rangle
$ in perturbation theory.
A simple calculation gives
\begin{eqnarray}
\langle \bar{\psi_i} \psi_i \rangle = &{}&\frac{3 m_i^3}{4\pi^2} \left[ 1 +
\ln (\frac{\mu^2}{m_i^2})
+ 2 \frac{\alpha_s}{\pi}
\left(
\ln^2(\frac{\mu^2}{m_i^2})
+ \frac{5}{3}
\ln (\frac{\mu^2}{m_i^2})
+ \frac{5}{3}
\right)
\right]
\nonumber
\\
&+&
\frac{ m_i \lambda^2}{4\pi^2}
\frac{\alpha_s}{\pi}
\left(
-5 + 6\ln \frac{\mu^2}{m_i^2}
\right)
{}.
\label{mqq}
\end{eqnarray}
Now in order to find the polarization operator $\Pi^{(1)}_V$ in the case of
normal non-ordered condensates one proceed by equating  $\Pi^{(1)}_{V,NO}$
to the combination
$\Pi^{(1)}_{V,NON} + \Pi^{(1)}_{V,con}$ with the condensate values taken
from Eq.~(\ref{mqq}). The result of the treatment indeed does not contain
any mass logs and reads
\begin{eqnarray}\label{Pi1m2GGv2}
(16\pi^2)\Pi^{(1)}_{V,NON} &=&
\Bigg{[}
\frac{20}{3}
+6 \frac{m_{-}^2}{Q^2}
-6 \frac{m_{+}^2}{Q^2}
+4 l_{\mu Q}
+6 \frac{m_{-}^2}{Q^2}l_{\mu Q}
\Bigg{]}
\nonumber\\
&{+}& \frac{\alpha_s}{\pi}
\Bigg{[}
\frac{55}{3}
-16 \,\zeta(3)
+\frac{107}{2} \frac{m_{-}^2}{Q^2}
-24 \,\zeta(3)\frac{m_{-}^2}{Q^2}
-16 \frac{m_{+}^2}{Q^2}\nonumber\\
&+& 4 l_{\mu Q}
+22 \frac{m_{-}^2}{Q^2}l_{\mu Q}
-12 \frac{m_{+}^2}{Q^2}l_{\mu Q}
+6 \frac{m_{-}^2}{Q^2}l_{\mu Q}^2
\Bigg{]}
\nonumber\\
&{+}& \frac{\alpha_s}{\pi}\frac{\lambda^2}{Q^2} \Bigg{[} -\frac{128}{3} +32
\,\zeta(3)
-\frac{76}{3} \frac{m_{-}^2}{Q^2}
+16 \,\zeta(3)\frac{m_{-}^2}{Q^2}\nonumber\\ &-& 8 \frac{m_{+}^2}{Q^2}
+12
\frac{m_{-}^2}{Q^2}l_{\mu Q}
-12 \frac{m_{+}^2}{Q^2}l_{\mu Q}
\Bigg{]}
{}.
\end{eqnarray}
\noindent
As it follows from Eq.~\re{mqq} the quark condensate obeys the non-trivial
RGE: \begin{equation}
\mu^2\frac{d}{d\mu^2} m_q
\langle
\overline{\psi}_{{\rm i}} \psi_{{\rm i}} \rangle = \frac{3 m_q m_i
\lambda^2}{2\pi^2}
\frac{\alpha_s}{\pi}
{},
\label{RGVEV}
\end{equation}
where
we have neglected all quartic quark mass terms. For further analysis, we
shall choose the subtraction point $\mu$ equal to the sum rule scale $M$,
as discussed in \cite{chk}.

In the light-quark case relevant to the
$\rho$-channels we can neglect the $m^2$ terms and $\Pi_{\rho}(M^2)$
simplifies greatly:

\begin{equation}
\Pi_{\rho}(M^2) =
(parton ~model)\cdot
\left(
1+
\left(\frac{{\alpha_s}}{\pi}\right)
\left[1 + \frac{\lambda^2}{M^2}
\left(
-\frac{32}{3} + 8\zeta(3)
\right)
\right]
\right)
\label{Bor.Pi1.male}
{},
\end{equation}
or, in numerical form,
\begin{equation}
\Pi_{\rho}(M^2) =
(parton ~model)\cdot
\left(
1
+\left(\frac{{\alpha_s}}{\pi}\right)
\left[
1 - 1.05\frac{\lambda^2}{M^2}
\right]
\right)
\label{Bor.Pi1.male.num}
{},
\end{equation}
This result was actually obtained earlier \cite{ball}. It is interesting
that terms of order $M^{-2}$ were looked for first \cite{narison} in
connection with proposal
(see first paper in Ref. \cite{yamawaki}) that UV renormalons can be
imitated by a
nonperturbative quark mass. In terms of the quark mass the correction is
substantially larger and is
$-{6m^2/ M^2}$ (see Eq. (\ref{Pi1m2GGv2})). The conclusion of the analysis
\cite{narison} was that the
data allow only for an {\it imaginary} quark mass and quite a small one:
\begin{equation} m_q^2\simeq
-(71-114)^2 ~\rm{MeV}^2 \end{equation} (further analysis favoured rather
lower end of the band \cite{narison1}). \\
Now, that we are interpreting possible $M^{-2}$ correction in terms of the
gluon mass, the conclusion \cite{narison} that a nonperturbative quark mass
could only be imaginary fits very well our scheme with a tachyonic gluon
mass. Moreover,
the value of the quark mass, say, $m_q^2=-0.01~\rm{GeV}^2$ is translated
into a much larger gluon
mass, $\lambda^2\approx -0.5~\rm{GeV}^2$. \\ To summarize our discussion of
the $\rho$-channel, we find an important independent indication that the
present data could be interpreted in terms of a tachyonic gluon mass:
\begin{equation}
\lambda^2 (1~\rm{GeV})\approx -(0.2-0.5)~\rm{GeV}^2~, \end{equation} where
we have
used $(\alpha_s/\pi)(1~\rm{GeV})\simeq 0.17\pm 0.02$. This result indicates
that, even relatively large masses , say, $\lambda^2\approx -0.5$ GeV$^2$
cannot be ruled out. A further improvement of the (axial-) vector spectral
function from $e^+e^-$ or/and $\tau$ decay data would improve the present
constraint on the eventual existence of such a term.
\subsection{Alternative estimate of $\lambda^2$ from the $\pi$-channel} We
shall be concerned with
the (pseudo)scalar two-point correlator: \begin{equation} \psi_{(5)}(Q^2)
\equiv i \int
d^4x ~e^{iqx} \ \langle 0\vert {\cal T} J_{(5)}(x) J^\dagger _{(5)}(0) \vert 0
\rangle ,
\end{equation}
built from current of the bilinear light quark fields: \begin{equation}
\label{scalar}
J_{(5)}(x)=(m_i+(-)m_j)\bar q_i(i\gamma_5)q_j, \end{equation} having the
quantum
numbers of the pseudoscalar $(O^{-+})$ $\pi$ or scalar
$(O^{++})~a^0/\delta$ mesons.
In the chiral limit ($m_u\simeq m_d=0$), the QCD expression of the
absorptive part
of the correlator reads:
\begin{equation}
\frac{1}{\pi}{\rm Im}\psi_{(5)}(s)\simeq
(m_i+(-)m_j)^2\frac{3}{8\pi^2}s\Bigg{[}1+
\left(\frac{{\alpha_s}}{\pi}\right)\left(
-2\log\left(\frac{s}{\mu^2}\right)
+\frac{17}{3}-4\frac{\lambda^2}{s}\right)\Bigg{]}~, \label{pis5:res}
\end{equation}
where one should
notice that the coefficient of the $\lambda^2$ term: \begin{equation}
b_{\pi}\approx
4b_{\rho}~.\end{equation}
\\
In order to see the effect of this term, we work with the Borel/Laplace sum
rule:
\begin{equation}\label{laplace}
{\cal L}(\tau)
= \int_{t_\leq}^{\infty} {dt}~\mbox{e}^{-t\tau} ~\frac{1}{\pi}~\mbox{Im}
\psi_{(5)}(t),
\end{equation}
{and} the corresponding ratio of moments: \begin{equation}\label{ratio}
{\cal R}(\tau) \equiv -\frac{d}{d\tau} \log {{\cal L}(\tau)},
\end{equation} where $t_\leq$
is the
hadronic threshold.
\begin{table*}[hbt]
\begin{center}
\caption{Ratio of
moments ${\cal R}(\tau)$ and value of $\lambda^2$} \begin{tabular}{c c c c c}
\hline
&&&& \\
$\tau\equiv 1/M^2$[GeV$^{-2}$]&${\cal
L}_{exp}/(2f_\pi^2m_\pi^4)$&$R_{exp}$&
$R^{\lambda^2=0}_{QCD}$&$-(\alpha_s/\pi)\lambda^2$[GeV]$^2$\\ &&&&\\ \hline
&&&& \\
1.4&1.20&$0.27\pm 0.06$&$0.66^{+0.50}_{-0.31}$&$0.09^{+0.08}_{-0.06}$ \\
&&&&\\ 1.2&1.29&$0.36\pm
0.07$&$0.79^{+0.42}_{-0.28}$&$0.10^{+0.09}_{-0.05}$\\ &&&&\\
1.0&1.42&$0.46\pm 0.10$&$0.94^{+0.34}_{-0.25}$&$0.11^{+0.07}_{-0.06}$\\
&&&&\\ 0.8&1.61&$0.58\pm
0.12$&$1.10^{+0.26}_{-0.21}$&$0.12\pm 0.06$\\ &&&&\\ 0.6&1.89&$0.73\pm
0.15$&$1.24^{+0.20}_{-0.17}$&$0.12\pm{0.06}$\\ &&&&\\ 0.4&2.29&$0.89\pm
0.16$&$1.33^{+0.15}_{-0.13}$&$0.10\pm 0.06$\\ &&&&\\ \hline 
&&&&\\
\end{tabular}
\end{center}
\end{table*}
\noindent
The QCD expression of the sum rule can be written as: \begin{eqnarray}
{\cal L}(\tau)&=&
\frac{3}{8\pi^2}\Big{[}\overline{m}_u(\tau)+\overline{m}_d(\tau)\Big{]}^2
\tau^{-
2} \Bigg{[}
1+\delta_0+(\delta_2+\delta_\lambda)\tau +\delta_4\tau^2+\delta_6\tau^3\Bigg{]}
\nonumber\\
-\frac{d{\cal
L}}{d\tau}&=&2\frac{3}{8\pi^2}\Big{[}\overline{m}_u(\tau)+
\overline{m}_d(\tau) \
Big{]}^2\tau^{-3}\Bigg{[}
1+\delta'_0+\frac{1}{2}\left(
\delta_2+\delta_\lambda\right)\tau-\frac{1}{2}\delta_6 \tau^3\Bigg{]}
\nonumber\\
{\cal R}(\tau)&\equiv&-\frac{d{\cal L}}{d\tau}\Big{/}{\cal L}~,
\end{eqnarray} where
\cite{becchi}--\cite{CHET}:
\begin{eqnarray}
\delta_0&=&4.821a_s+28.953a^2_s\nonumber\\ \delta'_0&=&6.821a_s+57.026a^2_s
\nonumber\\
\end{eqnarray}
while:
\begin{equation}
\delta_\lambda=-4a_s\lambda^2~.
\end{equation}
By keeping the leading linear and quadratic mass terms, one has
\cite{svz,snb}: \begin{eqnarray}
\delta_2&=&-2(\overline{m}^2_u+\overline{m}^2_d-\overline{m}_u
\overline{m}_d) \tau ,\nonumber\\
\delta_4&=&\frac{8\pi^2}{3}\left\{ \frac{1}{8\pi}\langle \alpha_s
G^2\rangle-\left(
m_d-\frac{m_u}{2}\right)
\langle \bar uu\rangle+(u\leftrightarrow d)\right\} ,\nonumber\\
\delta_6&=&k\frac{896}{27}\pi^3\alpha_s\langle \bar uu\rangle ^2 .
\end{eqnarray}
$\overline{m}_i(\tau)$
and $a_s\equiv \overline{\alpha}_s(\tau)/\pi$ are respectively the running
quark masses and QCD coupling.
The expression of the running coupling\index{running coupling} to two-loop
accuracy can be parametrized as ($\nu^2\equiv \tau^{-1}$): \begin{eqnarray}
a_s(\nu)\equiv
\left(\frac{\bar{\alpha_s}}{\pi}\right)
=a_s^{(0)}\Bigg\{ 1-a_s^{(0)}
\frac{\beta_2}{\beta_1}\log
\log{\frac{\nu^2}{\Lambda^2}}+{\cal{O}}(a_s^2)\Bigg\}~, \end{eqnarray}
with:
\begin{equation} a_s^{(0)}\equiv
\frac{1}{-\beta_1\log\left(\nu/\Lambda\right)}~, \end{equation} and
$\beta_i$ are the
${\cal{O}}(a_s^i)$ coefficients of the $\beta$ function\index{$\beta$
function} in the $\overline{\mbox{MS}}$-scheme for $n_f$ flavours, which,
for three
flavours, read:
\begin{equation}
\beta_1=-9/2~,~~~~~\beta_2=-8~.
\end{equation}
$\Lambda$ is a renormalization group
invariant\index{Renormalization Group
Invariant (RGI)}
scale but is
renormalization scheme\index{renormalization scheme} dependent. The
expression of the running quark mass in terms of the invariant
mass\index{invariant mass} $\hat{m}_i$ is \cite{snb}: \begin{eqnarray}
\overline{m}_i(\nu)&=&\hat{m}_i\left(
-\beta_1 a_s(\nu)\right)^{-\gamma_1/\beta_1}
\Bigg\{1+\frac{\beta_2}{\beta_1}\left( \frac{\gamma_1}{\beta_1}-
\frac{\gamma_2}{\beta_2}\right) a_s(\nu)+{\cal{O}}(a_s^2)\Bigg\}~
\end{eqnarray}
where
$\gamma_i$ are the ${\cal{O}}(a_s^i)$ coefficients of the quark-mass
anomalous dimension\index{anomalous dimension}. For three flavours, we
have: \begin{equation}
\gamma_1=2~,~~~~\gamma_2=91/12~.
\end{equation}
We shall use in our numerical analysis, the QCD parameters compiled in
\cite{snb}:
\begin{eqnarray}
\Lambda&=& (375\pm 50)~\rm{ MeV}\nonumber\\ \langle \alpha_s G^2\rangle&=&
(0.07\pm 0.01)~\rm {GeV}^4 \nonumber\\ \langle \bar
uu\rangle^{1/3}(1~\rm {GeV})&=& -(229\pm 9)~\rm{MeV} \nonumber\\
k&\simeq& 2-3\nonumber
\\
M^2_0&=&(0.8\pm 0.1)~\rm{GeV}^2
\end{eqnarray}
where $k$ indicates the deviation from the vacuum saturation assumption of
the four-quark condensates, while $M_0^2$ parametrize the mixed quark-gluon
condensate:
\begin{equation}
\langle\bar u G^a_{\mu\nu}\frac{\lambda_a}{2} u \rangle= M^2_0\langle\bar
uu\rangle~.
\end{equation}
One can notice that the ratio of moments ${\cal R}(\tau)$ is not affected
by the
quark masses to leading order of chiral symmetry breaking terms. Therefore,
we shall
use it for extracting $\lambda^2$.
We parametrize the pseudoscalar spectral function as usual by the pion pole
plus the
$3\pi$ contribution.The $3\pi$ threshold has been calculated using lowest
order chiral
perturbation theory, while two $\pi'$ resonances are introduced with the
following widths and
masses in units of MeV \cite{RAFAEL}: \begin{equation} M_1= 1300\pm 100
~~~\Gamma_1=400\pm 200
~~~~\rm{and}~~~~ M_2= 1770\pm 30 ~~~\Gamma_2=310\pm 50 \end{equation}
Including
finite width corrections, one obtains: \begin{equation} \frac{1}{\pi}{\rm
Im}\psi_5(s)=2f^2_\pi
m^4_\pi \Bigg{[}\delta(s-m^2_\pi)+ \theta(s-9m^2_\pi)
\frac{1}{(16\pi^2f_\pi^2)^2}\frac{s}{18}\rho_{had.}(s)\Bigg{]}
\end{equation} where
$\rho_{had.}(t)$
is given in \cite{RAFAEL}. We use, in our analysis, the largest range of
predictions given by
the two different parametrisations of the hadronic spectral function
proposed in \cite{RAFAEL},
which correspond to the two values $\xi=-0.23 +i~0.65$ (best duality with
QCD) and
$\xi=0.23+i~0.1$ (best fit to the experimental curve for the observed
$\pi'(1770)$ in hadronic
reactions) of the phenomenological interference complex parameter $\xi$
between the two $\pi'$
states. The result of this analysis corresponding to ${\cal R}_{exp}(\tau)$
is summarized in
Table 1, where we have also used $t_c\simeq 2.5$ GeV$^2$ as fixed from the
duality analysis in
\cite{RAFAEL}. Therefore, we can deduce to a good approximation, at the
$\tau-$stability region
($\tau\approx 0.8$ GeV$^{-2})$:
\begin{equation}
(\alpha_s/\pi)\lambda^2\simeq
\frac{1}{4}\Big{[} {\cal R}_{exp}-{\cal R}^{\lambda^2=0}_{QCD}\Big{]}
\simeq -(0.12\pm
0.06)~\rm{GeV}^2. \end{equation}
One can notice, by comparing this result with the one from the
$\rho$-channel, that, to leading order in $\lambda^2$, the two estimates
lead to
the consistent value:
\begin{equation}\label{lambda2}
(\alpha_s/\pi)\lambda^2\simeq
-(0.06-0.07)~\rm{GeV}^2~~~~\Longrightarrow~~~~~ \lambda^2 (\tau\approx
0.8~\rm{GeV}^{-2}) \simeq -(0.43\pm 0.09)~\rm{GeV}^2~. \end{equation}
One can also notice in Table 1 that the introduction of $\lambda$ has
enlarged the region of
QCD duality \cite{RAFAEL} or $\tau$ stability \cite{snb,snmass} to a lower
value of
$M^2\equiv 1/\tau$ of about
$M_\rho$. \\
To summarize, the pseudoscalar channel can be considered as a success for
the phenomenology with $\lambda^2\neq 0$.

\subsection{Effects of $\lambda^2$ on $\overline{m}_{u,d}$ and $\overline{m}_s$
from the (pseudo)scalar channels} Re-analysing the ${\cal L}$ sum rule by
the inclusion of this new
$\lambda$ term, one can also notice that the presence of such a term tends
to decrease only slightly
the prediction for the sum $(\overline{m}_{u}+\overline{m}_{d})$ of the
running light quark masses by
about:
\begin{equation}
\delta_{ud}\approx -5.6\%
\end{equation}
Applying this effect to the recent estimates in \cite{RAFAEL,snmass}, one
obtains:
\begin{equation}
(\overline{m}_u + \overline{m}_d)(1 ~{\rm{GeV}})=(11.3\pm
2.4)~{\rm{MeV}}~~~~ \Longrightarrow ~~~~ (\overline{m}_u +
\overline{m}_d)(2 ~\rm{GeV})=(8.6\pm 2.1)~\rm{MeV}~.
\end{equation}
For the estimate of the strange quark mass from the kaon channel, one also
obtains
a decrease of:
\begin{equation}
\delta_{us}\approx -5\%
\end{equation}
which is slightly lower than the one of the pion channel due to the partial
cancellation of
the $\lambda$ effect by the $m_s^2$ one.\\ \subsection{ $\lambda^2$ and the
value of $\alpha_s(M_\tau)$ from tau decays} A natural extension of the
analysis is to test the effect of $\lambda^2$ on the value of $\alpha_s$
extracted from tau-decays. For this purpose, we re-do the analysis from
$R_\tau$ given for the first in \cite{bnp}, by adding the new $\lambda^2$
term. The modified expression of the tau hadronic width is \cite{bnp}:
\begin{equation}
R_\tau= 3\left( (|V_{ud}|^2+|V_{us}|^2\right) S_{EW}^\tau\Big{\{}
1+\delta^\tau_{EW}+\delta^\tau_{PT}+ \delta^\tau_2+
\delta^\tau_{\lambda}+\delta^\tau_{NP}\Big{\}}. \end{equation} where:
$|V_{ud}|=0.9751\pm 0.0006$, $|V_{us}| =0.221\pm 0.003$ are the weak mixing
angles;
$S^\tau_{EW}=1.0194$ and $\delta^\tau_{EW} =0.0010$ are the electroweak
corrections;
$\delta_2^\tau$ is a small correction from the running light quark masses;
$\delta^\tau_{NP}$ are non-perturbative effects due to operators of
dimension $\geq$ 4.
\begin{table*}[hbt]
\begin{center}
\caption{Reduction of $\alpha_s(M_\tau)$ for different values of
$\lambda^2$} \begin{tabular}{c c}
\hline
& \\
$ -\lambda^2$ [GeV$^2$]&$-\delta^{\alpha_s}_\lambda [\%]$\\ &\\ \hline &\\
0.16& $4\pm 1$\\
&\\
0.25&$7\pm 1$ \\
&\\
0.34& $9\pm 1$\\
&\\
0.43& $11\pm 2$\\
&\\
0.52& $14\pm 2$\\
&\\
0.61& $16\pm 3$\\
&\\
\hline
\end{tabular}
\end{center}
\end{table*}
\noindent
The correction due to the new dimension-2 term is: \begin{equation}
\delta^\tau_\lambda= 2\frac{b_\rho}{M^2_\tau}. \end{equation} where, from
Eq. (\ref{Bor.Pi1.male.num}), $b_\rho=-1.05a_s\lambda^2$ and $\lambda^2$ is
given in Eq. (\ref{lambda2}). At fixed order of perturbation theory, one
has \cite{Gorishny91,bnp}: \begin{equation} \delta^\tau_{PT}=
a_s+5.2023a_s^2+26.366a_s^3+{\cal O}(a_s^4)~, \end{equation} where
we have
truncated the series at the complete computed coefficients. We use the
values of the QCD non-perturbative parameters given in \cite{narison1} and
the
experimental value \cite{aleph} $R_\tau=3.484\pm 0.024$. By comparing the
extracted value of
$\alpha_s(M_\tau)$ obtained using $\lambda^2=0$ and $\lambda^2\neq 0$, we
deduce in Table 2,
using the previous approximations, a reduction of the $\alpha_s(M_\tau)$-value
for different
values of $\lambda^2$. The error given in Table 2 reflects the uncertainty
on the eventual
scale-dependence of the quantity $a_s\lambda^2$ or $\lambda^2$. The quoted
central
value comes from the average between the result corresponding to the
freedom by taking
one of these two quantities as the input number in the analysis. \\

\noindent
$\bullet$ If we use the value of $\lambda^2$ as given in Eq. (\ref{lambda2}),
then, we get
to leading order in $\lambda^2$:
\begin{equation}
\delta^{\alpha_s}_\lambda \simeq -(11\pm 3)\%\label{thirteen}. \end{equation}
On the other hand, a
comparison, at the $\tau$-mass, of the value of $\alpha_s$ from the $Z$
or/and the world average
\cite{PDG} with the one from $\tau$ decays, where the latter includes
$\alpha_s^3$ corrections, gives a difference of about \cite{menke}:
\begin{equation}\label{five}
\delta^{\alpha_s}_\lambda \simeq -(5\pm 10)\%, \end{equation} which is of
the same
sign and compatible with the
correction in Eq. (\ref{thirteen}). One should also have in mind that there
are obvious limitations on the accuracy of the estimate
in Eq. (\ref{thirteen}), due to the no control so far over terms
proportional to $\lambda^2$
and higher powers of $\alpha_s(M^2_{\tau})$, including the anomalous
dimension of the tachyon
mass. These unknown effects might be taken into account by enlarging the
error in Eq.
(\ref{thirteen}) by about a factor 2.\\

\noindent
$\bullet$ Inversely, one can improve
the previous determinations of
$\lambda^2$ from the
$\rho$ and
$\pi$ channels, either by adding, in the different experimental fits
\cite{aleph} of the
moments of the $\tau$-decays
\cite{bnp,ledi}, the new $\lambda^2$ term with the extra (compared with
previous experimental
fits) constraint that the sign of its contribution is unambiguously fixed,
or by attributing the
eventual small deviation of the values of $\alpha_s$ from $\tau$ and from
e.g. $Z-$decays as
being due to this new term. If one proceeds in the later way, and use the
result in Eq.
(\ref{five}), one would obtain from Table 2: \begin{equation}
\lambda^2 (M_\tau) \simeq
-(0.2\pm 0.4)~\rm{GeV}^2~,
\end{equation}
which is less conclusive than the result in Eq. (\ref{lambda2}). \\

\noindent
$\bullet$ Since the introduction of the gluon mass still presumably gives a
better control over
$1/M^2_{\tau}$ terms which have been introduced so far ad hoc and are the
major sources
of theoretical uncertainties \cite{sntau}--\cite{neubert}, one expects,
within the present
approach, an improvement of the theoretical errors in the determination of
$\alpha_s(M_\tau)$ by
about a factor 2.
\\

\noindent
$\bullet$ Furthermore, by comparing our previous results with some other
approaches which resum
the higher
order terms of the perturbative series \cite{sntau}--\cite{neubert}, the
effect of the
$1/M^2$ term becomes manifest
when one extends
the analysis of the tau hadronic widths to a lower hypothetical
$\tau$-lepton mass
\cite{bnp,narison1,aleph}. We have therefore checked that until
$M_\tau\approx 1.3$ GeV, the
agreement of the prediction with the recent data \cite{aleph} remains quite
good.
\\

\noindent
$\bullet$ Finally, one should notice that a na\"{\i}ve redefinition of
$\alpha_s$ by
a coupling which contains implicitly a $\lambda^2/M^2$ term, in the
expression of
the physical hadronic width $R_\tau$, but not on the level of the propagator in
Eq. (\ref{simple}), will lead to inconsistencies.

\subsection{Effects of $\lambda^2$ on the value of $\overline{m}_s$ from
$e^+e^-$
and $\tau$ decays} In the determinations of $m_s$ based on the ratio
\cite{snmass}
or on the difference \cite{chen} of the non-strange and strange quark
channels,
the
leading terms which are  flavour independent will not contribute.
Taking the leading
term in
$m^2_s\lambda^2$, which can be obtained from the previous expression of
$\Pi^{(1)}_V$, one can deduce that the presence of the $\lambda$ term
implies a slight decrease of the value of $m_s$ of about (1--3)\%, which
is, however, smaller
than the quoted errors (15\%) in its determination.\\ In the case of the
determinations of
$m_s$ from the alone
$\Delta S=1$ part of the inclusive $\tau$ decay rate \cite{pich},
one can check, after
using the resummed series given there, and the
expressions of the different corrections in the massless case given in the
previous
subsection, that the presence of $\lambda$ tends to increase by about
(8--9)\% the
value of $m_s$ obtained from this method. Again, this effect is still
smaller than
the large uncertainties (about 30\%) of this determination from this method. \\
However, the effect of $\lambda$ tend to improve the agreement betwen the two
different determinations.
\subsection{The quest of the scalar $a^0/\delta$ and $K^*_0$-channels}
Consideration of the
correlator of the scalar $J_{\delta}$ current (or of its strange analogue
$J_{K^*_0})$, where \begin{equation}
J_{\delta}={1\over \sqrt{2}}\left(\bar{u}u-\bar{d}d\right) \end{equation}
brings most
interesting problems and
challenges to the phenomenology with $\lambda^2\neq 0$. \\
Indeed, in the
limit of massless quarks, the (ordinary) QCD expressions of the pion and
scalar sum rules only differ for the contributions of the dimension-six (or
more) condensates, where here: \begin{equation}
\delta_6=-k\frac{1408}{81}\pi^3\alpha_s\langle \bar uu\rangle ^2,
\end{equation} such that
the strength of $M^2_{crit}$ or the optimization scale of about 2 GeV$^2$
\cite{snb}
is obviously the
same in both channels, i.e. much larger than the one of the $\rho$-channel.
However, one should also
notice that, in this case of ordinary OPE ($\lambda^2=0$ and neglect of the
instanton
effects), the scalar sum rules can reproduce within a good accuracy the
experimental mass of
the $a^0/\delta$ meson, while the predicted decay constant leads to a width
consistent with the present
data \cite{snb}. In the same way, the estimate of the running light quark
masses from the scalar sum
rules
\cite{paver,snb,JAMIN,CHET} within the ordinary OPE gives a prediction
consistent with the analysis
from other methods \cite{PDG} or from the $e^+e^-$ \cite{snmass}, $\tau$
decay data \cite{pich} or a
global sum rule extraction of the light quark condensate \cite{dosch}. \\
It is also obvious that $M^2_{crit}$ associated with $\lambda^2\neq 0$ is
the same
in the pion and $\delta
$-channels:
\begin{equation}
M^2_{crit} (\pi-channel)~\approx~ M^2_{crit}(\delta-channel).
\end{equation} What may be even a more
drastic prediction is that the sign of the leading $M^{-2}$ correction is
the same in the two channels. Moreover, a detailed analysis of the
$a_0/\delta$ sum rule in the
case $\lambda^2\neq 0$ gives acceptable phenomenological results, and moves
the scale of optimization
to lower values of $M^2_{crit}$ like the case of the pion. \\
On the other hand,
direct instantons predict the same $M^2_{crit}$ in the both channels but
the opposite signs for the leading correction \cite{shuryak}. Thus, the
scalar channel might be the right place to discuss in more details the
choice between the direct
instantons and the $M^{-2}$ corrections. \\
On pure theoretical grounds,
both direct instantons and nonperturbative $M^{-2}$ corrections seem to be
indispensable parts of the QCD phenomenology. Indeed, instantons are
necessary to resolve
the $\eta^{'}$-problem. The nonperturbative $M^{-2}$ terms are necessary to
match the $M^{-2}$
uncertainty of the perturbative series due to the UV renormalons. \\
All
these qualitative arguments cannot fix, however, which correction (if any
of the two) becomes
important first at large $M^{2}$. The signatures of the direct instantons
and of the tachyonic mass
are different in the $\pi$- and $\delta$-channels. Namely, the tachyonic
mass gives the same-sign
correction while direct instantons result in opposite signs for the
deviations from the asymptotic
freedom. The existing lattice data \cite{chu} indicate, to our mind, that
it is rather a mixture of
the two mechanisms which works. Indeed, the corrections in the two channels
are rather of opposite
signs but the correction in the pion channel is substantially stronger. In
the $\delta$-channel the
correction is much smaller than it should be if the instanton-model
parameters are normalized to the
$\pi$-channel data.
\\
Further data with better accuracy could be helpful. For example, it is not
ruled out that in the $\delta$-channel the correction is first positive
because of the gluon mass and
becomes negative at lower $M^2$ because of the effect of the direct instantons.

\subsection{The gluonia channels}
We shall first be concerned with the two-point correlator: \begin{equation}
\Pi_G(Q^2)~
\equiv~ i \int d^4x ~e^{iqx} \ \langle 0\vert {\cal T} J_G(x) \left( J_G(0)
\right)^\dagger\vert 0 \rangle
\end{equation}
associated to the scalar gluonium current: \begin{equation}
J_G={\beta(\alpha_s)\over \alpha_s}(G^a_{\mu\nu})^2. \end{equation} Its
evaluation leads to:
\begin{equation}
\label{GG_scalar_res}
\Pi_G(M^2) = (parton~ model)
\left(1-{3\lambda^2\over M^2}+...\right) {}. \label{scalarGG:res}
\end{equation}
Thus, one can expect that the $\lambda^2$ correction in this channel is
relatively
much larger since it is not proportional to an extra power of $\alpha_s$.
With $\lambda^2 \approx-0.5$ GeV$^2$ we obtain, by using the same 10 per
cent convention as in section 3:
\begin{equation}\label{mcglue}
M^2_{crit}(0^+-gluonium)~\approx~15 ~\rm{GeV}^2 \end{equation} in amusing
agreement with the independent estimate in Eq. (\ref{glue2}). \\

\noindent
Exactly the same phenomenon is observed for the case of the two point
correlator
of the pseudoscalar gluonium currents: the relative strength of the
$\lambda^2/M^2$ term added
to the parton result coincides with that for the scalar gluonium displayed
in Eq.~\re{GG_scalar_res}.
\\

\noindent
However, one should notice
that a more quantitative analysis based on $\tau$-stability of the
corresponding (pseudo)scalar
sum rules \cite{nv}--\cite{shore} leads to a lower value of $M^2_{crit}\approx
(3-5)$ GeV$^2$, but still much
larger than the scale of the $\rho$ meson.\\

\noindent
Let us consider now the case of the tensor gluonium with the correlator:
\begin{eqnarray} &{}&
\psi^T_{\mu\nu\rho\sigma}(q)\equiv i\int d^4x~ e^{iqx} \langle 0|
{\cal T}
\theta^g_{\mu\nu}(x)
\theta^g_{\rho\sigma}(0)^\dagger|0\rangle\nonumber \nonumber \\ &=&
\psi^T_4 \left(
q_\mu q_\nu q_\rho q_\sigma
-\frac{q^2}4
(q_\mu q_\nu g_{\rho\sigma} + q_\rho q_\sigma g_{\mu\nu})
+\frac{q^4}{16}(g_{\mu\nu}g_{\rho\sigma} ) \right) \nonumber \\ &+&\psi^T_2
\left(
\frac{q^2}{4} g_{\mu\nu} g_{\rho\sigma} - q_\mu q_\nu g_{\rho\sigma} -
q_\rho q_\sigma g_{\mu\nu}
+ q_{\mu}q_{\sigma} g_{\nu\rho}
+ q_{\nu}q_{\sigma} g_{\mu\rho}
+ q_{\mu}q_{\rho} g_{\nu\sigma}
+ q_{\nu}q_{\rho} g_{\mu\sigma}
\right)
\nonumber
\\
&+&\psi^T_0
\left(
g_{\mu\sigma} g_{\nu\rho}
+
g_{\mu\rho}g_{\nu\sigma}
-
\frac{{1}}{{2}} g_{\mu\nu} g_{\rho\sigma} \right) {}\, ,
\label{tensor_def1} \end{eqnarray}
where
\begin{equation}
\label{tensor_def}
\theta^g_{\mu\nu}= -G_{\mu}^{\alpha}G_{\nu\alpha} + \frac{1}{4} g_{\mu\nu}
G_{{\alpha_\beta}} G^{\alpha\beta} {}\, . \end{equation}
A direct calculation gives the following results for the structure
functions $\psi^T_i$ and their respective Borel/Laplace transforms:
\begin{eqnarray}
\label{psi4}
\pi^2 \psi^T_4 &=&
\frac{l_{\mu Q}}{15}
+
\frac{17}{450}
-
\lambda^2 \frac{1}{3 Q^2}
{}
=\!=\!\Longrightarrow
\frac{1}{15}
\left(
1 - 5 \frac{\lambda^2}{M^2}
\right)
{}\, ,
\\
\label{psi2}
\pi^2 \psi^T_2
&=&
\frac{Q^2 l_{\mu Q}}{20}
+\frac{9 Q^2}{200}
+\lambda^2
\left(
\frac{l_{\mu Q}}{6}
-\frac{2}{9}
\right)
=\!=\!\Longrightarrow
-\frac{M^2}{20}
\left(
1 - \frac{10}{3}\frac{\lambda^2}{M^2}
\right)
{}\, ,
\\
\label{psi0}
\pi^2 \psi^T_0 &=&
\frac{Q^4 l_{\mu Q}}{20}
+\frac{9 Q^4}{200}
+\lambda^2 Q^2
\left(
\frac{l_{\mu Q}}{4}
-\frac{1}{12}
\right)
=\!=\!\Longrightarrow
\frac{M^4}{20}
\left(
1 - \frac{5}{2}\frac{\lambda^2}{M^2}
\right)
{}\, .
\end{eqnarray}
If, instead of considering $\theta_{\mu\nu}^g$, we would introduce the
total energy-momentum tensor of interacting quarks and gluons
$\theta_{\mu\nu}$, then various functions components of
$\psi_{\mu\nu\rho\sigma}$
are related to each other because of the energy-momentum conservation.
Indeed, requiring that
\[
\psi^T_{{\mu\nu\rho\sigma}} q_\mu \equiv 0 \] we immediately obtain: \[
\psi^T_2 = \frac{3}{4} Q^2 \psi^T_4
\ \ \ \mbox{and} \ \ \
\psi^T_0 = \frac{3}{4}Q^4 \psi^T_4
{}\, ,
\]
and, a consequence, the following representation of the function in Eq.
\re{tensor_def1}:
\begin{eqnarray}
\psi^T_{\mu\nu\rho\sigma}(q)
&=&
\left(
\eta_{\mu\rho}\eta_{\nu\sigma}+\eta_{\mu\sigma}\eta_{\nu\rho} - \frac{2}{3}
\eta_{\mu\nu}\eta_{\rho\sigma}\right) \psi^T(Q^2), \label{temsor_def1}
\end{eqnarray}
where:
\begin{equation}\label{psiT}
\psi^T(Q^2) \equiv Q^4 \frac{3}{4}\psi^T_4(Q^2), \ \ \ \eta_{\mu\nu}\equiv
g_{\mu\nu}-\frac{q_\mu q_\nu}{q^2} {}.
\end{equation}
To compare the sum rules based on $\lambda^2\neq 0$ with the existing ones
with $\lambda^2=0$ \cite{nv,narison3}, one should have in mind that the results in
\cite{nv} come from the sum
rules corresponding to the correlator $\psi^T$ defined in Eq. (\ref{psiT}).
Literally,
these sum rules are not affected by the new $\lambda^2$ terms because, in
terms of the
imaginary parts of the structure functions, the $\lambda^2$ correction
exhibited in Eq.
(\ref{psi4}) is proportional to $\lambda^2\delta(s) $ which vanishes once
multiplied by an
extra power of $s$. However, further analysis of once more subtracted sum
rules as well as of the $\tau-$stability \cite{nv} would reveal now a large
mass scale.

\noindent
To summarize,
we conclude, from Eqs (\ref{psi4}--\ref{psiT}) above, that the
Borel/Laplace transforms of
the functions $\psi^T_i(i=0,2,4)$ as well as $\psi^T$ will have a scale
analogue to the one of the (pseudo)scalar gluonium channels. This feature
might reveal
that all gluonia channels have an universal scale , in distinction from the
instanton model
\cite{shuryak} where the tensor-gluonium channel is expected to be
insensitive to the largest
scale exhibited in the spin-0 gluonium channel. Measurements of the
corresponding correlators
could provide therefore an interesting test of the theory.
\subsection{Heavy quarks: $\bar{Q}Q$
and $\bar{Q}q$-channels.} Introduction of the gluon mass would result in a
substantial change in
the QCD phenomenology of heavy quark interactions. However, we expect that
it would be rather a
reshuffle of the parameters than a decisive test. Indeed, there are
successful description of the
quarkonia states with the linear potential in Eq. (\ref{potential})
extrapolated down to $r=0$,
see, e.g.,
\cite{tye}. We plan to come back to these points in a future publication.
\section{Summary and conclusions}
To summarize our discussions of the various channels, the introduction of
the tachyonic gluon mass allows to explain in a very simple and unified way
the variety of
scales of the violation of the asymptotic freedom at moderate $Q^2$ in
various channels. More
specifically, the phenomenology based on the introduction of
$\lambda^2\simeq -0.5$ GeV$^2$
leads to the following successful predictions: \\

\noindent
$\bullet$ The correct sign and order of magnitude of the $M^{-2}$
correction in the $\rho$-channel.\\

\noindent
$\bullet$ The new $M^{-2}$ correction in the pion channel breaks the
asymptotic freedom in
this channel at the mass scale $M^2_{crit}$ which is about factor of 4
higher than in the
$\rho$-meson channel. This scale falls close to the value of $M^2_{crit}$
determined
independently from the values of $f_{\pi}$ and quark masses. The sign of
the correction due to
the tachyonic mass is also the one which is needed to bring QCD in
agreement with phenomenology.
\\

\noindent
$\bullet$ A natural explanation of $M^2_{crit}$ in the scalar-gluonium
channel (Eq.
(\ref{mcglue}))
which is much larger than $M^2_{crit}$ in the $\rho$-channel and which was
found
independently from a
low energy theorem (Eq. (\ref{glue2})).
\\

\noindent
$\bullet$ The account for the $M^{-2}$ correction lowers by $(11\pm 3)\%$
the value of $\alpha_s(M_{\tau})$ determined from the $\tau$-decays, and
may improve the accuracy of its determination by a factor of about 2. It
also decreases by about 5$\%$ the value of the $u,~d$ and $s$ running quark
masses from the pseudoscalar channels. The presence of this term slightly
decreases by about (1--3) \% the value of $m_s$ obtained
from $e^+e^-$ \cite{snmass} and $\tau$ decay
data \cite{chen}. It increases by about (8--9)\% the results in the
in \cite{pich} from the individual $\Delta S=1$ inclusive $\tau$ decay
channel. This
effect, then, improves the agreement between the two different
determinations. \\

\noindent
To overview the logic of our analysis,
we were motivated to introduce a tachyonic gluon mass by the data on the
$Q\bar{Q}$ potential
at short distances (see section 2). In this case $\lambda^2\neq 0$ brings
in a linear term $kr$ at short distances which is, however, only a small
correction to
the Coulombic potential. Any precise measurement of this correction would
require a subtraction
of large perturbative contributions and as a result there are no error bars
available on the
value of $k$ at short distances. Thus, the idea was to embed this
(hypothetical) linear term in
the potential into a relativistic framework and consider further
consequences from this
extension. Hence, the introduction of the tachyonic gluon mass. This
extension allowed to
analyse the current correlators. As a result, the bounds on the $\lambda^2$
narrow
substantially. In particular, $\lambda^2\approx-$1 GeV$^2$ which would
naively correspond to the
known value of $k$ does not pass phenomenological hurdles, which is not so
surprising. What is
much more amusing, the value $\lambda^2\approx-$0.5 GeV$^2$ fits the data
well (see above).\\
Most remarkably, the introduction of $\lambda^2\neq 0$ brought large
qualitative effects,
namely a variety of the mass scales $M^2_{crit}$ where the asymptotic
freedom is broken at various channels. Various $M^2_{crit}$ may differ by a
factor of about 20.
Further crucial checks of the phenomenology based on $\lambda^2\neq 0$
could be provided by
measurements of the correlators on the lattice. The $a^0/\delta$-scalar
channel and gluonia
channels appear most interesting from this point of view. \\ However, the
extension of the
standard QCD sum rules by introducing $\lambda^2\neq 0$ cannot resolve all
the phenomenological
problems. Namely, the $\eta^{'}$-problem cannot be solved in this way and
asks for introduction
of instantons or instanton-like configurations. The existing data on
lattice simulations might
indicate that in the $a^0/\delta$-scalar channels, direct instantons become
important at about
the same $M^2$ as the effects of the gluon mass do. Further measurements of
the current
correlators on the lattice could provide checks of the phenomenology
explored in the present
paper.
\section*{Acknowledgements}
The work by K.G. Chetyrkin has been
supported by DFG under Contract Ku 502/8-1 and INTAS under Contract
INTAS-93-744-ext.\\
S. Narison wishes to thank M. Davier and S. Menke for useful
communications, and A. Pich
for comments. \\
V.I.
Zakharov gratefully acknowledges hospitality of Centre National de la
Recherche Scientifique
(CNRS) during his stay at the University of Montpellier where part of this
work was done. He is
also thankful to R. Akhoury, G. Bali, L. Stodolsky for valuable
discussions. 

\newpage


\begin{figure}
\begin{center}
\epsfxsize=4.0cm
\leavevmode
\epsffile{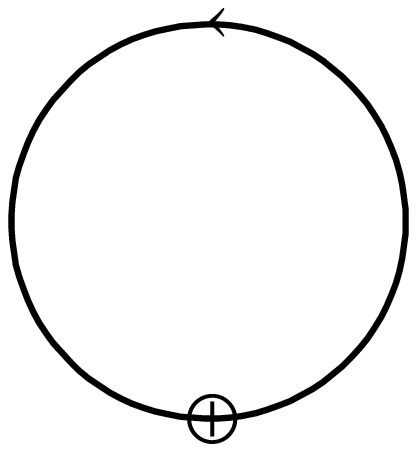}
\epsfxsize=4.0cm
\leavevmode
\epsffile{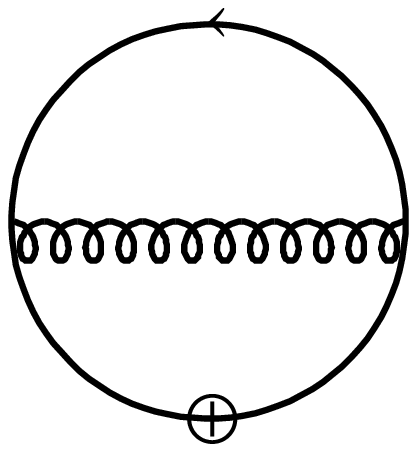}
\epsfxsize=4.0cm
\caption{\label{diag}Diagrams giving rise to nonzero vacuum expectation
value of the
operator $\overline{\psi} \psi$ in the lowest orders of perturbation
theory.} \end{center}
\end{figure}

\end{document}